\documentclass[]{article}
\usepackage{epsf,a4}
\begin{document}
\title{Protein dynamics with off-lattice\\ Monte Carlo moves}
\author{Daniel Hoffmann\thanks{hoffmann@Chemie.FU-Berlin.DE}\quad and 
\quad Ernst-Walter Knapp\thanks{knapp@Chemie.FU-Berlin.DE}\\
Institut f\"ur Kristallographie, Fachbereich Chemie,\\ 
Freie Universit\"at Berlin, D-14195 Berlin, Germany}
\date{Scheduled tentatively for Phys.Rev.E issue of 1 March 1996}
\maketitle

\begin{abstract}
A Monte Carlo method for dynamics simulation of all-atom protein
models is introduced, to reach long times not accessible to
conventional molecular dynamics. The considered degrees of freedom are
the dihedrals at C$_\alpha$-atoms. Two Monte Carlo moves are used:
single rotations about torsion axes, and cooperative rotations in
windows of amide planes, changing the conformation globally and
locally, respectively. For local moves Jacobians are used to obtain an
unbiased distribution of dihedrals. A molecular dynamics energy
function adapted to the protein model is employed. A polypeptide is
folded into native-like structures by local but not by global moves.
\end{abstract}

\vspace{0.5cm}

\vspace{0.5cm}
Two of the great challenges in theoretical biophysics are the
elucidation of the structure function dynamics relationship of
proteins \cite{general}, and the protein folding problem
\cite{Kol1986,Bal1990,Rad1992,theory}. Functional processes in
proteins are often slow compared to picosecond and subpicosecond
dynamics of vibrational degrees of freedom, which are easily
accessible by conventional methods of computer simulation. The
dynamics of protein folding with a typical time scale from
microseconds to seconds is even much slower.

The conventional method of computer simulation of protein dynamics is
based on solving Newton's equations of motion
in Cartesian coordinates \cite{Lev1982,Bro1983}. This
approach is very time consuming for the following reasons:
1.  The number of non-bonded atom pair interactions is very large, so
     that most of the CPU time is spent on their evaluation.
2.  It is necessary to follow the fast intramolecular vibrations in
     all details, which requires an elementary step of propagation in
     time of typically 1~fs.

There are two important strategies \cite{Gun1991} to improve this
situation:
1.  Reducing the number of non-bonded pair interactions by
      combining groups of atoms in single interaction centers. The
      drawback is a loss of detail in the description of the
      protein.
2.  Increasing the elementary time step by eliminating stiff degrees
freedom with small amplitudes of motion. These are bond lengths and
bond angles, which here are not interesting in their own right,
although they may serve as ``lubricant'' for large conformational
changes \cite{Kar1986}.

Lattice models, and protein models using virtual bonds instead of
amide planes are very efficient combinations of both time saving
techniques, but they suffer from poor energetic and conformational
resolution \cite{simplified}.

In the present approach we use an all-atom protein model where bond
lengths and bond angles are fixed and the amide plane is kept planar,
so that the only degrees of freedom of the protein backbone are the
dihedral angle pairs at the C$_\alpha$-atoms \cite{Kna1993b}
(Fig.~\ref{FigWindow}). New conformations are generated by moves of
two different types.  A simple move (SM) is a rotation about a torsion
axis. It leads to large displacements of atoms far away from this
axis. Thus in a globular protein SM's often result in structures with
overlapping atoms, which are energetically unacceptable. A Window move
(WM) is a cooperative rotation in a small window of consecutive amide
planes \cite{Kna1993b} (Fig.~\ref{FigWindow}).  WM's have been
employed to prove that a tripeptide can not be cyclic \cite{Go1970}
and to generate loop conformations \cite{Bru1985}. The finding of WM's
is equivalent to the problem of inverse kinematics for serial
manipulators, where the hand of a robot must be oriented and
positioned in a specific place \cite{robot}. The cooperative motion of
several amide planes can be described as a diffusion process. In this
sense a chain of WM's can be interpreted as dynamics evolution.

In a WM the changes of the dihedrals are subject to six constraints
which guarantee that the atomic positions of the two parts of the
polypeptide outside the window remain fixed \cite{Kna1993b}.  A window
can consist of any number of amide planes larger than one. Here only
windows of three amide planes are considered, corresponding to eight
degrees of freedom (Fig.~\ref{FigWindow}). The first steps in a WM are
the pre-rotations, where increments to the dihedrals at one
C$_\alpha$-atom in a window are chosen arbitrarily from a given
interval $[-\zeta,+\zeta]$. Possible values of the other six dihedrals
are then determined by finding a root of the constraining
conditions. For the present geometry of the protein backbone the
maximum number of roots found was twelve. On the average a complete
set of roots is obtained in less than 10~ms CPU time (SGI R4000),
which is negligible in comparison to the time spent on the evaluation
of the non-bonded energy.

The distribution of dihedral angles of protein backbone conformations
generated by WM's deviates from a uniform distribution by more than
30\% due to the window constraints. More specifically it turns out
that this bias concerns four dihedrals $\varphi_k$ out of eight in a
window. A corresponding reduced set of four constraints $c_l$ can be
used as generalized coordinates which trivially fulfill the window
constraints.  For the $i$th root of the constraint equations, the
Jacobian
\begin{equation}
\label{EqJacobian}
J_i=\left|\frac{\partial(\varphi_1,\varphi_2,\varphi_3,\varphi_4)}
               {\partial(c_1,c_2,c_3,c_4)}\right|_{(i)}
\end{equation}
accounts for the change of phase space volume when transforming from
the $c_l$ to the $\varphi_k$. The fact that different roots have
unequal $J_i$ is the reason for the bias. The bias is eliminated as
described in the following. After the pre-rotations have been carried
out in a WM, the window constraints are solved and one obtains $N_{\rm
new}$ new window conformations. All $N_{\rm new}$ conformations belong
to a single set of $c_l$ but have different values of $\varphi_k$ and
$J_i$. A weighting among the $N_{\rm new}$ conformations according to
their phase space volumes can be reached by using probabilities $p_i$
that obey $p_i\propto J_i, i=1,2,...,N_{\rm new}$. A correct weighting
between old and new conformations in the window requires also
knowledge of the $N_{\rm old}$ conformations, which are obtained by
solving the window constraints without applying
pre-rotations. Consequently a proper weighting is achieved by randomly
selecting one of all old and new conformations according to the
probability
\begin{equation}
\label{EqProbability}
p_i=J_i\left(\sum_{k=1}^N J_k\right)^{-1},\; i=1,2,...,N
\end{equation}
where $N = N_{\rm old}+N_{\rm new}$. If there are no solutions for the
applied pre-rotations, the old conformation is retained. The necessity
of the correct weighting of solutions has often been ignored and was
first recognized by Dodd {\em et al.} \cite{Dod1993} in the context of
polymer dynamics. The present selection scheme is more efficient than
the rejection algorithm used in Ref.~\cite{Dod1993}. In
the two windows at the ends of the polypeptide chain, the dihedral
angle changes are randomly chosen from $[-\zeta,+\zeta]$ without
further constraints. The final decision on acceptance or rejection of
a WM is left to the Metropolis criterion \cite{Met1953} with a
suitable energy function. With $N_w$ different window positions
possible, a chain of $N_w$ WM's with randomly chosen window positions
is called a ``scan''. For SM's the term scan is used in the same
spirit, with $N_w$ corresponding to the total number of torsion axes.

Since the protein backbone model has reduced flexibility, a
conventional MD energy function cannot be used without
modifications. Atomic clashes may occur between atoms separated by
only a few torsional degrees of freedom. In a protein model with full
flexibility this could be avoided by bond angle bending. In this work
these problems are circumvented by letting adjacent amide planes
interact only by a two-dimensional potential of the two dihedrals at
the C$_\alpha$-atom connecting the amide planes, which is generated
beforehand with CHARMM as described in appendix 2 of
ref.~\cite{Bro1983}. These two-dimensional potentials are residue
specific; here only those of glycine and alanine are needed. For all
other atom pairs, which are separated by at least four torsional
degrees of freedom the CHARMM interactions are used.

In this contribution the folding of an
$\alpha$-helix-turn-$\alpha$-helix (HTH) structure is simulated. The
HTH motif is guided by the structure of ROP \cite{Ban1987}, an
$\alpha$-helical hairpin of 56 residues. In the simulations, only 26
residues are considered, corresponding to residues 18-43 of ROP. The
$\alpha$-helical parts are modeled by alanine (A) residues. It is
known from MD simulations, that polyalanine forms stable
$\alpha$-helices in vacuo \cite{Dag1991}. A hydrophobic attraction of
the helices is mimicked by specifically increasing the well-depth of
the C$_\beta$-C$_\beta$ Lennard-Jones interaction from 0.181 to
2.0~kcal/mol between some of the residues (X), corresponding to those
residues in ROP which line the interface of the two helices. The
well-depth of the X-X interaction is motivated by the measured free
energies of transfer of hydrophobic amino acids from water to
non-polar solvents (corresponding to the hydrophobic core of
proteins), which for leucin or phenylalanin are about 2.0~kcal/mol
\cite{Cho1974}. The turn region is modeled by five glycines (G), which
are the most flexible residues. In total the sequence of the model
protein reads AXAAXAAAXXGGGGGXXAAAXAAAXA. N- and C-terminus are
blocked with the neutral groups acetamide and N-methyl-amide,
respectively, to avoid strong electrostatic interactions. By simulated
annealing and subsequent energy minimization the ROP model structure
is adapted to the sequence and energy function of the HTH model
yielding a root mean square deviation (RMSD) of 1.35~{\AA} for the
backbone. The energy of this annealed reference structure (RS) $E_{\rm
RS} = -1215$~kcal/mol, and its radius of gyration with respect to the
C$_\alpha$-atoms $\gamma_{\rm RS} = 7.1$~{\AA} correlate well with the
values obtained by MC dynamics with WM's as discussed below.

Four trajectories using WM's and four using SM's have been produced
(WM$i$ and SM$i, i=1,2,3,4$). All trajectories start from a
$\beta$-strand conformation, where the chain is almost extended,
$E=-1008$~kcal/mol, and $\gamma=24.7$ \AA. The temperature is 450~K,
which in a series of test simulations was found to be low enough for
formation of stable conformations but high enough for fast
isomerization. In WM simulations dihedrals are allowed to change per
move by at most $\zeta=20^\circ$. The global character of SM's makes
atomic clashes more likely, hence we choose $\zeta=10^\circ$ for
SM's. In spite of its smaller $\zeta$ a SM can change the conformation
far more than a WM. For a protein of $N$ monomers, the average number
of non-bonded energy terms affected by a SM scan and a WM scan is
proportional to $N^3/3 + O(N^2)$ and $3N^2 + O(N)$, respectively. For
a 26-mer the average CPU time per SM scan is therefore about 2.8 times
that per WM scan. This difference has been approximately considered by
choosing the WM trajectories twice as long as the SM trajectories.

With the exception of SM1, in all SM trajectories the polypeptide
collapses within the first few thousand scans into compact random coil
type conformations. This is reflected in an abrupt decrease of
$\gamma$ from 25~{\AA} to less than 7~{\AA} (Fig.~\ref{FigRog}). After
the collapse $\gamma$ jumps erratically by up to 0.5~{\AA}. The
conformations are thus on the average more compact than the RS. They
have low secondary structure content with only a few isolated helix
turns. The energy $E$ falls very steeply by about 100 kcal/mol within
the first 1000 SM scans of SM2-SM4 (Fig.  \ref{FigE}). This is due to
X-X-contacts and hydrogen bonds, which form instantly but arbitrarily
between sequentially distant monomers as the polypeptide becomes
kinked. A slower decrease of $E$ follows over $1\times10^5 -
2\times10^5$ SM scans. Afterwards only smaller conformational
rearrangements take place, accompanied by energy fluctuations about
mean values of -1175~kcal/mol (SM2,SM3) and -1150 kcal/mol (SM4). Even
after energy minimization none of the conformations with lowest energy
of each SM trajectory fulfills $E<E_{\rm RS}+5$~kcal/mol
(Fig.~\ref{FigConf}).

SM1 leads to the conformations with the lowest $E$, the highest helix
content and largest $\gamma$ of all SM simulations. Within the first
$2\times10^4$ scans the central helix (Fig.~\ref{FigConf}) in SM1
forms and $E$ falls considerably. The growing of this helix slows down
the collapse to a denser conformation. Unfortunately the helix
encompasses just the glycines which are known to have low helix
propensity.

SM's require a relative mobility of chain ends. Therefore SM
trajectories are often trapped in quasi-cyclical conformations with
chain ends linked by strong hydrogen bonds or X-X-contacts.  The
dropping of acceptance probabilities from more than 0.6 within the
first SM scans to typically 0.30 when the conformation has become
quasi-cyclic reflects this feature.

In the WM trajectories the folding takes a very different
path. Starting from the termini two helices grow towards the center of
the polypeptide, which allows $E$ to decrease from -1008~kcal/mol to
about -1175~kcal/mol within the first $10^5$ scans. It was reported
that in MD simulations a 13-mer polyalanine requires several hundred
picoseconds to form an $\alpha$-helix \cite{Bro1989}. In test
simulations of 13-alanine with WM's helices were formed in several
thousand scans. Equating a MD time step with a WM scan, the CPU time
for helix formation is 100 times larger for MD than for MC. In WM4 the
two helices join after $10^5$ scans in the middle of the chain, which
is then a single long helix with average values for $E$ and $\gamma$
of -1175~kcal/mol and 11.7~{\AA}, respectively. The helix frays at the
ends and bends but remains stable. In WM1-WM3 the two helices form a
HTH-motif. The turn develops within the first $5\times10^5$ scans as
the polypeptide is kinked in the glycine region. The two helices are
forced into an anti-parallel alignment by the X-X attractions, leading
to a further decrease of $E$ to a mean value of -1180~kcal/mol
(Fig.~\ref{FigE}). The value of $\gamma$ drops, too, and fluctuates
finally at about $\gamma_{\rm RS}=7.1$~{\AA}
(Fig.~\ref{FigRog}). After the formation of the HTH-motifs the
conformations continue to rearrange, because they have either
imperfect pairing of X-residues (WM2,WM3) or helices with left-handed
turns (WM1,WM3). Nevertheless the trajectories WM1-WM3 contain
conformations, which after minimization have $E<E_{\rm RS}$
(Fig.~\ref{FigConf}), in particular for WM1, $E=E_{\rm
RS}-8$~kcal/mol, and the RMSD to RS is 1.9~{\AA}.  The acceptance
probability for WM's lies at 0.40 and is a product of a probability of
0.66 for the generation of a new conformation and a probability of
0.60 for the acceptance in the Metropolis algorithm.

The Monte Carlo simulations of a model protein have demonstrated that
WM's, which produce gradual and local conformational changes, first
lead to a quick formation of secondary (helices) and then a slower
development of tertiary (HTH) structure. The same pattern is seen in
real folding processes \cite{Rad1992,Kuw1987}. Conformational
reorganization decreases as the simulations proceed, but continues
till the end.  SM simulations tend to become trapped after a fast
initial collapse into compact but disordered conformations, because
the global conformational changes of SM's often require to break
several hydrogen bonds and hydrophobic contacts. None of the SM
trajectories correlates with RS, whereas three out of four WM
trajectories come close to RS. Up to now it was thought that it is not
feasible to address the protein folding problem for detailed protein
models. The present work shows that with WM's this problem can be
tackled successfully. The method will also be useful for the
simulation of polymer models in general.

The authors are grateful to Fredo Sartori for providing the code of
the energy function and to the Deutsche Forschungsgemeinschaft for
financial support.

\newpage
\section*{Figures}

{
\begin{figure}[htb]
        \leavevmode
        \epsfxsize=100mm
        \centering
        \epsffile[34 500 561 718]{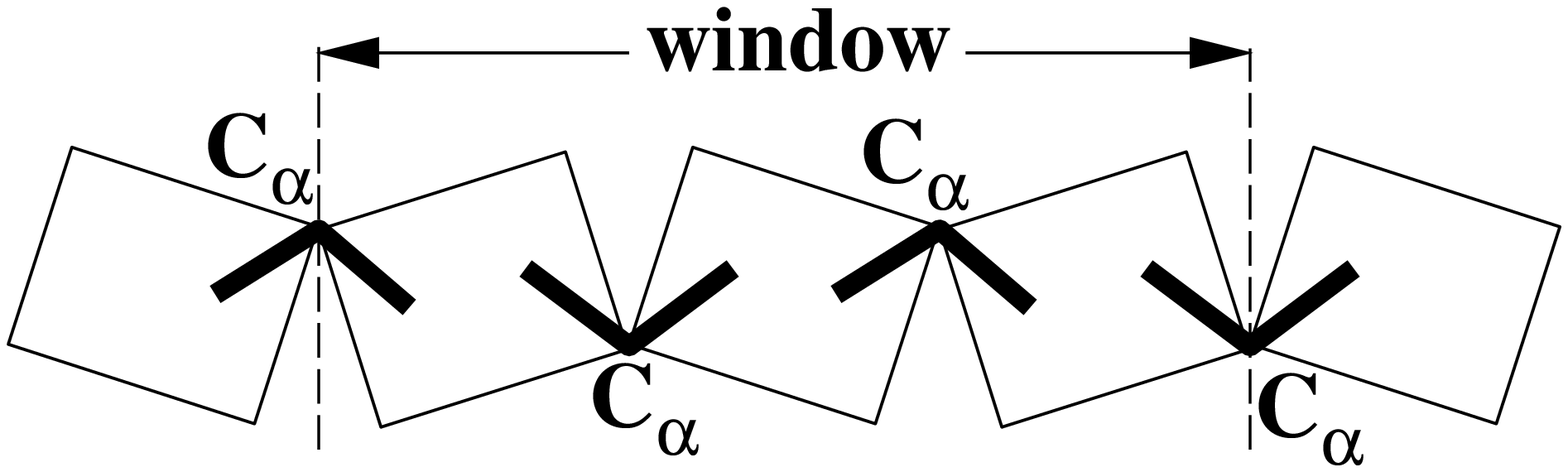}
\caption{Sketch of a window of three amide planes within the protein
backbone. The torsion axes corresponding to the eight degrees of
freedom are indicated by solid bars.}
\label{FigWindow}
\end{figure}
}

{
\begin{figure}[htb]
        \leavevmode
        \epsfxsize=100mm
        \centering
        \epsffile[136 300 448 683]{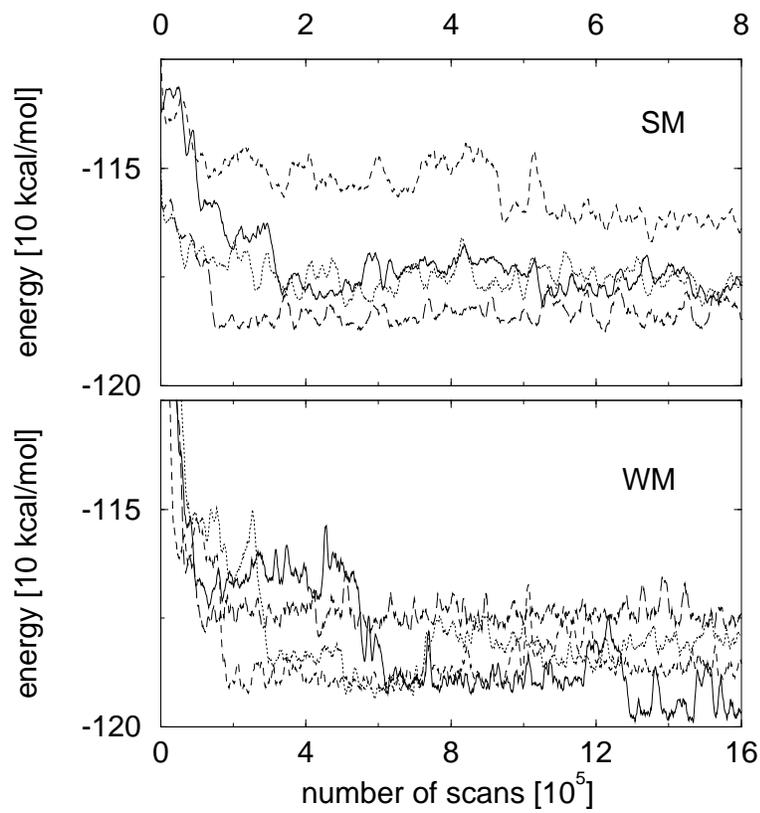}
\caption{Energy traces. The upper part shows SM1
(long dashed), SM2 (dotted), SM3 (solid) and SM4 (dashed). The lower
part shows WM1 (solid), WM2 (dashed), WM3 (dotted) and WM4 (long
dashed). Abscissas are scaled differently. Data are
smoothed by a running average over $10^4$ scans.}
\label{FigE}
\end{figure}
}

{
\begin{figure}[htb]
        \leavevmode
        \epsfxsize=100mm
        \centering
        \epsffile[152 300 448 683]{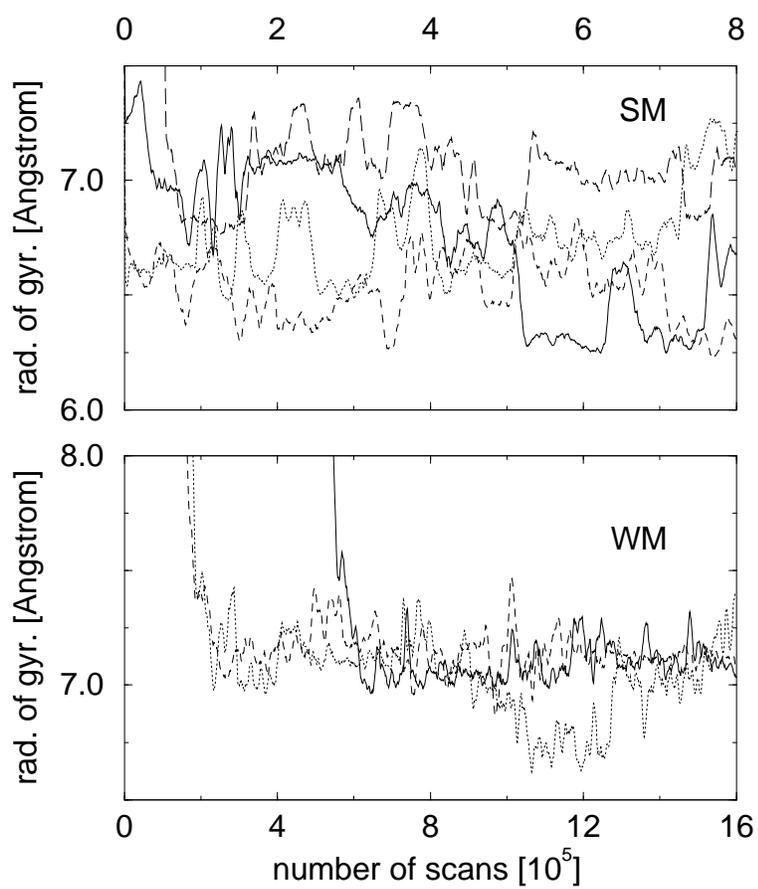}
\caption{Radius of gyration $\gamma$ with respect to the
C$_\alpha$-atoms. For line types and smoothing see
Fig.~\ref{FigE}. Different ordinates are used for top and bottom. WM4
is not shown (see text).}
\label{FigRog}
\end{figure}
}

{
\begin{figure}[htb]
        \leavevmode
        \epsfxsize=100mm
        \centering
        \epsffile[171 350 412 683]{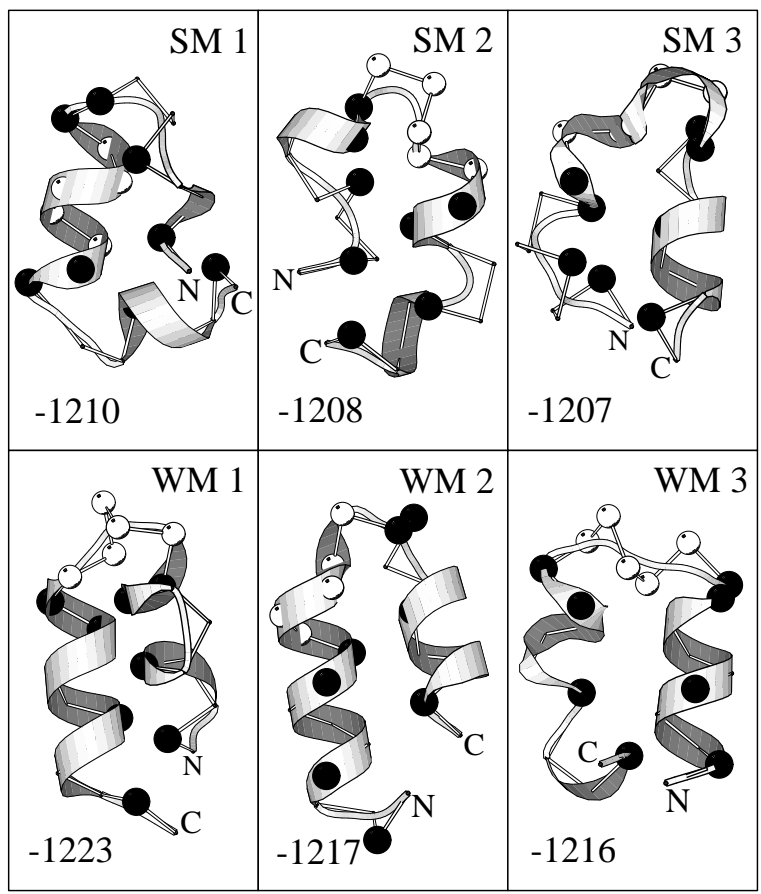}
\caption{Conformations with minimum energies. From each trajectory the
conformation with minimum energy was selected and energy minimized
(inserted: energies in kcal/mol). C$_\alpha$-atoms are connected by
sticks and those of G (X) are drawn as open (filled) circles. Wide
ribbons are helix turns [22]. SM4 and WM4 are not shown (see text).}
\label{FigConf}
\end{figure}
}
\end{document}